\documentclass[conference]{IEEEtran}
\usepackage{amsmath,graphicx,url,times}
\usepackage[bookmarks=false]{hyperref}
\usepackage[usenames, dvipsnames]{color}
\usepackage{multirow}
\ifCLASSINFOpdf
\else
\fi
\begin{document}
%
\title{Convolutional Gated Recurrent Neural Network Incorporating Spatial Features for Audio Tagging}

\author{\IEEEauthorblockN{Yong Xu~~~~~~Qiuqiang Kong~~~~~~Qiang Huang~~~~~~Wenwu Wang~~~~~~Mark D. Plumbley}
	\IEEEauthorblockA{Center for Vision, Speech and Sigal Processing
		University of Surrey, Guildford, UK\\
		Email: \{yong.xu, q.kong, q.huang, w.wang, m.plumbley\}@surrey.ac.uk}}

\maketitle

\begin{abstract}
Environmental audio tagging is a newly proposed task to predict the presence or absence of a specific audio event in a chunk. Deep neural network (DNN) based methods have been successfully adopted for predicting the audio tags in the domestic audio scene. In this paper, we propose to use a convolutional neural network (CNN) to extract robust features from mel-filter banks (MFBs), spectrograms or even raw waveforms for audio tagging. Gated recurrent unit (GRU) based recurrent neural networks (RNNs) are then cascaded to model the long-term temporal structure of the audio signal. To complement the input information, an auxiliary CNN is designed to learn on the spatial features of stereo recordings. We evaluate our proposed methods on Task 4 (audio tagging) of the Detection and Classification of Acoustic Scenes and Events 2016 (DCASE 2016) challenge. Compared with our recent DNN-based method, the proposed structure can reduce the equal error rate (EER) from 0.13 to 0.11 on the development set. The spatial features can further reduce the EER to 0.10. The performance of the end-to-end learning on raw waveforms is also comparable. Finally, on the evaluation set, we get the state-of-the-art performance with 0.12 EER while the performance of the best existing system is 0.15 EER.
\end{abstract}


%
\IEEEpeerreviewmaketitle

\section{Introduction}
Audio tagging (AT) aims at putting one or several tags on a sound clip. The tags are the sound events that occur in the audio clip, for example, ``speech", ``television", ``percussion", ``bird singing", and so on. Audio tagging has many applications in areas such as information retrieval \cite{wold1996content}, sound classification \cite{giannoulis2013detection} and recommendation system \cite{cano2005content}. 

Many frequency domain audio features such as mel-frequency cepstrum coefficients (MFCCs) \cite{molau2001computing}, Mel filter banks feature (MFBs) \cite{nadeu2001time} and spectrogram \cite{kingsbury1998robust} have been used for speech recognition related tasks \cite{hinton2012deep} for many years. However, it is unclear how these features perform on the non-speech audio processing tasks. Recently MFCCs and the MFBs were compared on the audio tagging task \cite{xu2016fully} and the MFBs can get better performance in the framework of deep neural networks. The spectrogram has been suggested to be better than the MFBs in the sound event detection task \cite{espi2015exploiting}, but has not yet been investigated in the audio tagging task.

Besides the frequency domain audio features, processing sound from the raw time domain waveforms, has attracted a lot of attentions recently \cite{sainath2015learning, tuske2014acoustic, golik2015convolutional}. However, most of this works are for speech recognition related tasks; there are few works investigating raw waveforms for environmental audio analysis. For common signal processing steps, the short time Fourier transform (STFT) is always adopted to transform raw waveforms into frequency domain features using a set of Fourier basis. Recent research \cite{sainath2015learning} suggests that the Fourier basis sets may not be optimal and better basis sets can be learned from raw waveforms directly using a large set of audio data. To learn the basis automatically, convolutional neural network (CNN) is applied on the raw waveforms which is similar to CNN processing on the pixels of the image \cite{krizhevsky2012imagenet}. Processing raw waveforms has seen many promising results on speech recognition \cite{sainath2015learning} and generating speech and music \cite{oord2016wavenet}, with less research in non-speech sound processing.

Most audio tagging systems \cite{xu2016fully, lidy2016cqt, cakirdomestic, yundiscriminative} use mono channel recordings, or simply average the multi-channels as the input signal. However, using this kind of merging strategy disregards the spatial information of the stereo audio. This is likely to decrease recognition accuracy because the intensity and phase of the sound received from different channels are different. For example, kitchen sound and television sound from different directions will have different intensities on different channels, depending on the direction of the sources. Multi-channel signals contain spatial information which could be used to help to distinguish different sound sources. Spatial features have been demonstrated to improve results in scene classification \cite{eghbal2016cp} and sound event detection \cite{adavannesound}. However, there is little work using multi-channel information for the audio tagging task. 

Our main contribution in this paper includes three parts. First, we show experimental results on different features including MFBs and spectrogram as well as the raw waveforms on the audio tagging task of the DCASE 2016 challenge. Second, we propose a convolutional gated recurrent neural network (CGRNN) which is the combination of the CNN and the gated recurrent unit (GRU) to process non-speech sounds. Third, the spatial features are incorporated in the hidden layer to utilize the location information. The work is organized as follows: in Section \ref{sec:cnn_gru}, the proposed CGRNN is presented for audio tagging. In section \ref{sec:spatial_fea}, the spatial features will be illustrated and incorporated into the proposed method. The experimental setup and results are shown in Section \ref{sec:exp_setup} and Section \ref{sec:results}. Section \ref{sec:conclusion} summarizes the work and foresees the future work. 

\section{Convolutional Gated Recurrent Network for audio tagging}
\label{sec:cnn_gru}
Neural networks have several types of structures: the most common one is the deep feed-forward neural network. Another popular structure is the convolutional neural network (CNN), which is widely used in image classification \cite{ciresan2011flexible, krizhevsky2012imagenet}. CNNs can extract robust features from pixel-level values for images \cite{krizhevsky2012imagenet} or raw waveforms for speech signals \cite{sainath2015learning}. Recurrent neural network is the third structure which is often used for sequence modeling, such as language models \cite{mikolov2010recurrent} and speech recognition \cite{graves2013speech}. In this section, we will introduce the convolutional neural network and the recurrent neural network with gated recurrent units.

\begin{figure}[t]
	\centering
	\centerline{\includegraphics[width=\columnwidth]{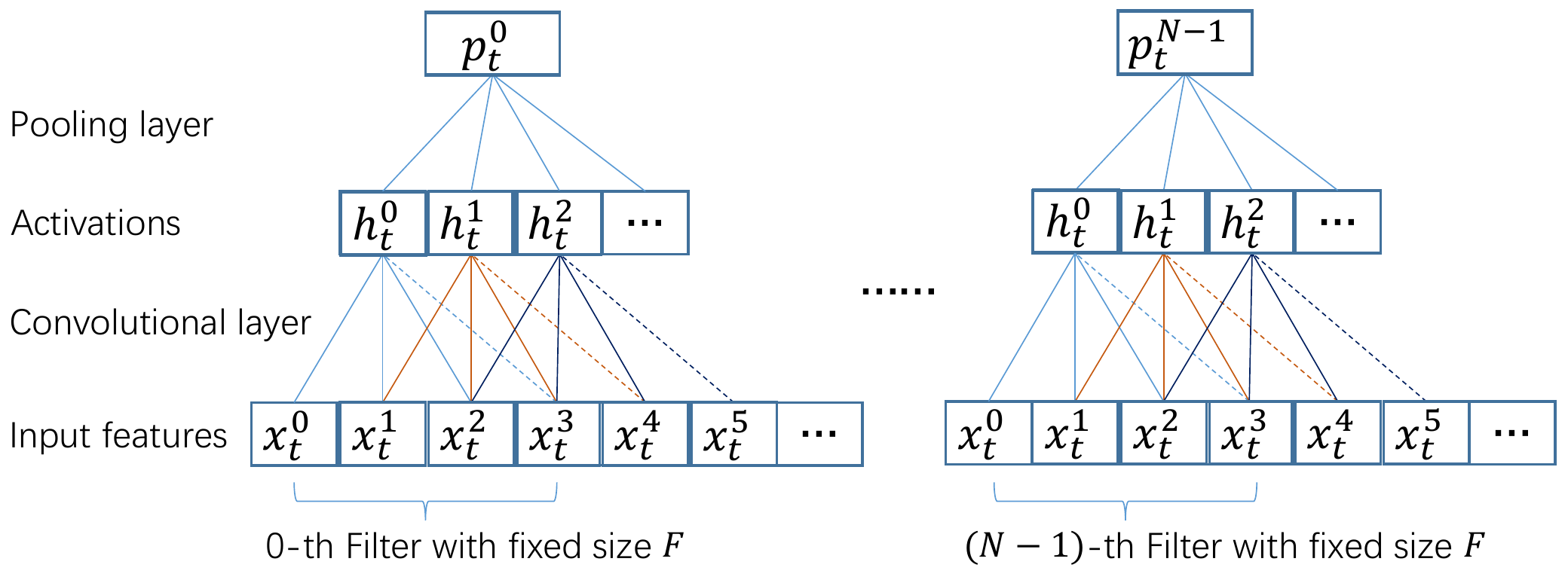}}
	\caption{The structure of the one-dimension CNN which consists of one convolutional layer and one max-pooling layer. $N$ filters with a fixed size $F$ are convolved with the one dimensional signal to get outputs $p_t^i\{i=0,\cdots,(N-1)\}$. ${x}_t^i$ means the $i$-th dimension feature of the current frame.}
	\label{fig:cnn}
\end{figure}

\subsection{One dimension convolutional neural network}
Audio or speech signals are one dimensional. Fig. \ref{fig:cnn} shows the structure of a one-dimension CNN which consists of one convolutional layer and one max-pooling layer. $N$ filters with a fixed size $F$ are convolved with the one dimensional signal to get outputs $p_i^t\{i=0,\cdots,(N-1)\}$. Given that the dimension of the input features was $M$, the activation $\textit{h}$ of the convolutional layer would have $(M-F+1)$ values. The max-pooling size is also $(M-F+1)$ which means each filter will give one output value. This is similar to speech recognition work \cite{sainath2015learning} where CNN has been used to extract features from the raw waveform signal. The way of each filter producing one value can also be explained as a global pooling layer which is a structural regularizer that explicitly enforces feature maps to be confidence maps of meaningful feature channels \cite{lin2013network}. So $N$ activations are obtained as the robust features from the basic features. As for the input feature size $M$, a short time window, e.g., 32 ms, was fed into the CNN each time. The long-term pattern will be learned by the following recurrent neural network. As for the filter size or kernel size, a large receptive field is set considering that only one convolutional layer is designed in this work. For example, $F=400$ and $M=512$ are set in \cite{sainath2015learning}. If the input feature was raw waveforms, each filter of the CNN was actually learned as a finite impulse response (FIR) filter \cite{golik2015convolutional}. If the spectrograms or mel-filter banks were fed into the CNN, the filtering was operated on the frequency domain \cite{abdel2013exploring} to reduce the frequency variants, such as the same audio pattern but with different pitches.

\subsection{Gated recurrent unit based RNN}
\begin{figure}[t]
	\centering
	\centerline{\includegraphics[width=\columnwidth]{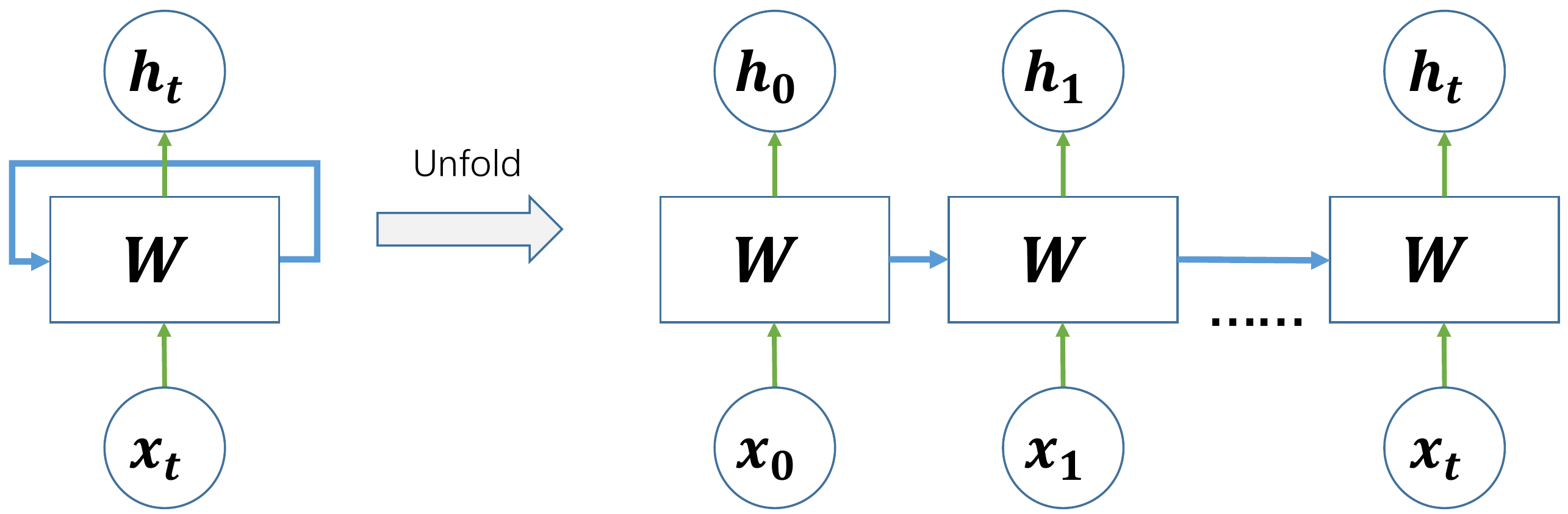}}
	\caption{The structure of the simple recurrent neural network and its unfolded version. The current activation $\textbf{h}_t$ is determined by the current input $\textbf{x}_t$ and the previous activation $\textbf{h}_{t-1}$.} 
	\label{fig:rnn}
\end{figure}

Recurrent neural networks have recently shown promising results in speech recognition \cite{graves2013speech}. Fig. \ref{fig:rnn} shows the basic idea of the RNN. The current activation $\textbf{h}_t$ is determined by the current input $\textbf{x}_t$ and the previous activation $\textbf{h}_{t-1}$. RNN with the capability to learn the long-term pattern is superior to a feed-forward DNN, because a feed-forward DNN is designed that the input contextual features each time are independent. The hidden activations of RNN are formulated as:
\begin{equation}
	{\textbf{h}}_t=\varphi(\textbf{W}^h\textbf{x}_t+{\textbf{R}^h\textbf{h}_{t-1}}+\textbf{b}^h)
	\label{eq:simple_rnn}
\end{equation}
However, a simple recurrent neural network with the recurrent connection only on the hidden layer is difficult to train due to the well-known vanishing gradient or exploding gradient problems \cite{pascanu2013difficulty}. The long short-term memory (LSTM) structure \cite{hochreiter1997long} was proposed to overcome this problem by introducing input gate, forget gate, output gate and cell state to control the information stream through the time. The fundamental idea of the LSTM is memory cell which maintains its state through time \cite{wu2016investigating}.

As an alternative structure to the LSTM, the gated recurrent unit (GRU) was proposed in \cite{cho2014learning}. The GRU was demonstrated to be better than LSTM in some tasks \cite{chung2014empirical}, and is formulated as follows \cite{wu2016investigating}:

\begin{equation}
\textbf{r}_t=\delta(\textbf{W}^r\textbf{x}_t+\textbf{R}^r\textbf{h}_{t-1}+\textbf{b}^r)
\end{equation} 
\begin{equation}
\textbf{z}_t=\delta(\textbf{W}^z\textbf{x}_t+\textbf{R}^z\textbf{h}_{t-1}+\textbf{b}^z)
\end{equation} 
\begin{equation}
\tilde{\textbf{h}}_t=\varphi(\textbf{W}^h\textbf{x}_t+\textbf{r}_t\odot{(\textbf{R}^h\textbf{h}_{t-1})}+\textbf{b}^h)
\label{eq:reset_gate}
\end{equation}
\begin{equation}
\textbf{h}_t=\textbf{z}_t\odot{\textbf{h}_{t-1}}+(1-\textbf{z}_t)\odot{\tilde{\textbf{h}}_t}
\label{eq:update_gate}
\end{equation} 
where $\textbf{h}_t$, $\textbf{r}_t$ and $\textbf{z}_t$ are hidden activations, reset gate values and update gate values at frame $t$, respectively. The weights applied to the input and recurrent hidden units are denoted as $\textbf{W}^*$ and $\textbf{R}^*$, respectively. The biases are represented by $\textbf{b}^*$. The functions $\delta(\cdot)$ and $\varphi(\cdot)$ are the sigmoid and tangent activation functions. Compared to the LSTM, there is no separate memory cell in the GRU. The GRU also does not have an output gate, and combines the input and forget gates into an update gate $\textbf{z}_t$ to balance between the previous activation $\textbf{h}_{t-1}$ and the update activation $\tilde{\textbf{h}}_t$ shown in Eq. (\ref{eq:update_gate}). The reset gate $\textbf{r}_t$ can decide whether or not to forget the previous activation (shown in Eq. (\ref{eq:reset_gate})). $\odot$ in Eq. (\ref{eq:reset_gate}) and (\ref{eq:update_gate}) represents the element-wise multiplication.

\subsection{Convolutional Gated Recurrent Network for audio tagging}
\begin{figure}[t]
	\centering
	\centerline{\includegraphics[width=\columnwidth]{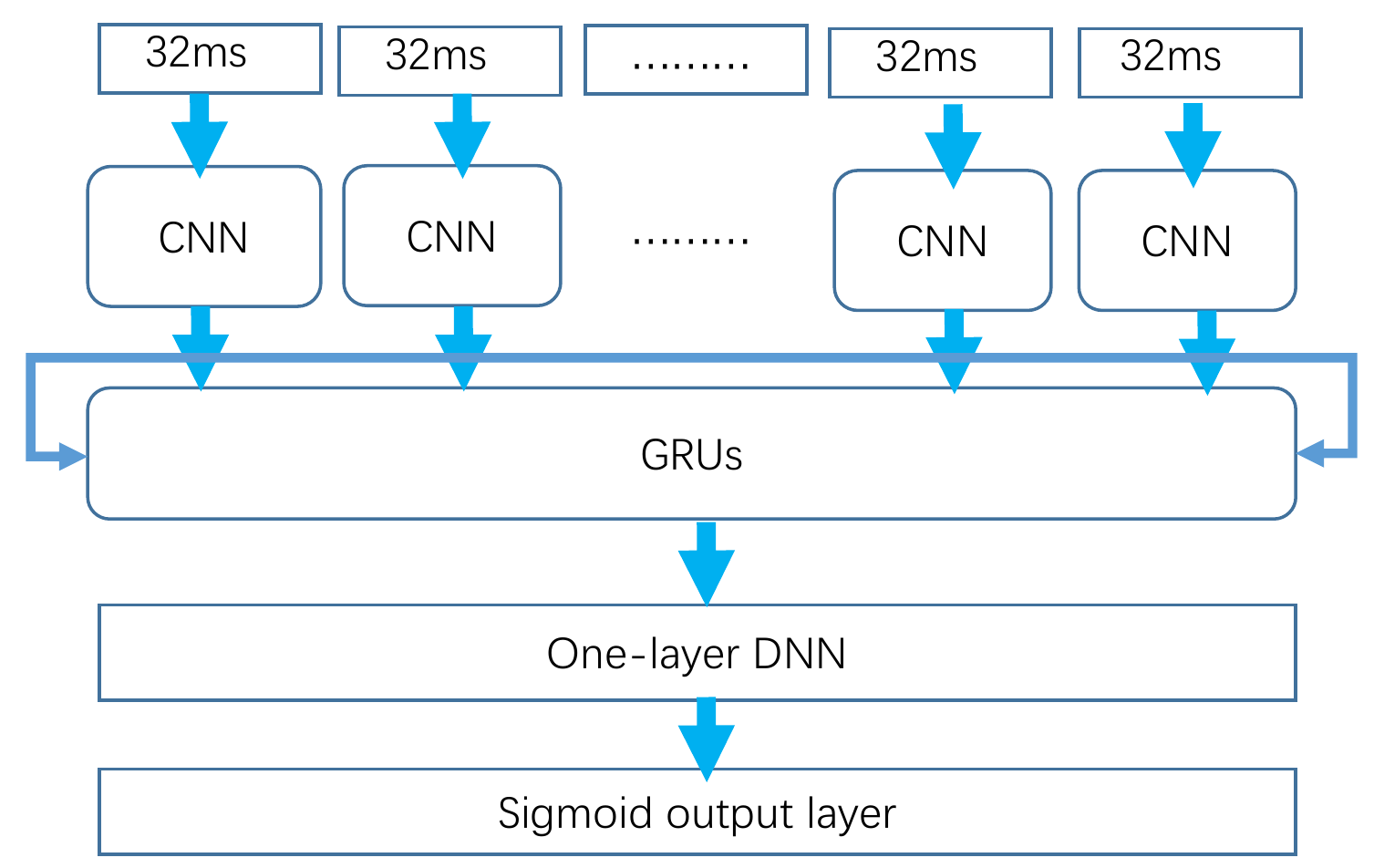}}
	\caption{The framework of convolutional gated recurrent neural network for audio tagging.}
	\label{fig:cnn_rnn_at}
\end{figure}

Fig. \ref{fig:cnn_rnn_at} shows the framework of a convolutional gated recurrent neural network for audio tagging. The CNN is regarded as the feature extractor along the short window (e.g., 32ms) from the basic features, e.g., MFBs, spectrograms or raw waveforms. Then the robust features extracted are fed into the GRU-RNN to learn the long-term audio patterns. For the audio tagging task, there is a lot of background noise, and acoustic events may occur repeatedly and randomly along the whole chunk (without knowing the specific frame locations). The CNN can help to extract robust features against the background noise by the max-pooling operation, especially for the raw waveforms. Since the label of the audio tagging task is at the chunk-level rather than the frame-level, a large number of frames of the context were fed into the whole framework. The GRU-RNN can select related information from the long-term context for each audio event. To also utilize the future information, a bi-directional GRU-RNN is designed in this work. Finally the output of GRU-RNN is mapped to the posterior of the target audio events through one feed-forward neural layer, with sigmoid output activation function. This framework is flexible enough to be applied to any kinds of features, especially for raw waveforms. Raw waveforms have lots of values, which leads to a high dimension problem. However the proposed CNN can learn on short windows like the short-time Fourier transform (STFT) process, and the FFT-like basis sets or even mel-like filters can be learned for raw waveforms. Finally, one-layer feed-forward DNN gets the final sequence of GRUs to predict the posterior of tags.

Binary cross-entropy is used as the loss function in our work, since it was demonstrated to be better than the mean squared error in \cite{xu2016fully} for labels with zero or one values. The loss can be defined as:
\begin{equation}
E=-\sum_{n=1}^{N}\|\textbf{T}_{n}\text{log}\hat{\textbf{T}}_{n}+(1-\textbf{T}_{n})\text{log}(1-\hat{\textbf{T}}_n)\|
\label{eq:DNNerrors_bce}
\end{equation}
\begin{equation}
\hat{\textbf{T}}_{n}=(1+\text{exp}(-\textbf{O}))^{-1}
\label{eq:DNN_hidden_sigmoid}
\end{equation}
where $E$ is the binary cross-entropy, $\hat{\textbf{T}}_n$ and $\textbf{T}_{n}$ denote the estimated and reference tag vector at sample index $n$, respectively. The DNN linear output is defined as $\textbf{O}$ before the sigmoid activation function is applied. Adam \cite{kingma2014adam} is used as the stochastic optimization method. 

\section{Spatial features incorporated for audio tagging}
\label{sec:spatial_fea}
\begin{figure}[t]
	\centering
	\centerline{\includegraphics[width=2in]{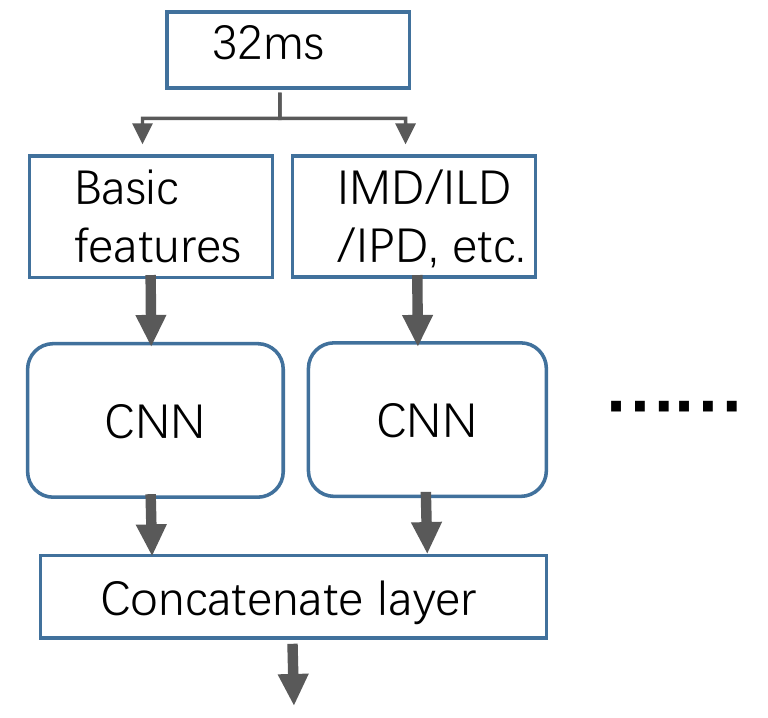}}
	\caption{The structure of incorporating the spatial features (IMD/ILD/IPD, etc.) using an additional CNN. Then the activations learned from the basic features and the activations learned from the spatial features are concatenated to be fed into the GRU-RNN shown in Fig. \ref{fig:cnn_rnn_at}.}
	\label{fig:cnn_imd}
\end{figure}
Spatial features can often offer additional cues to help to solve signal processing problems. Many spatial features can be used for audio tagging, such as interaural phase differences or interaural time differences (IPD or ITD) \cite{blauert1997spatial}, interaural level differences (ILD) \cite{blauert1997spatial}. The recordings of the audio tagging task of DCASE 2016 challenge are recorded in the home scenes. There are some TV, child speech, adult speech audio events. The spatial features potentially give additional information to analyze the content of the audio, e.g., recognizing the TV audio event by knowing the specific direction of the TV sound. The IPD and ILD are defined as:
\begin{equation}
ILD(t,k)=20\text{log}_{10}\left|\frac{X_{\text{left}}(t,k)}{X_{\text{right}}(t,k)}\right| \\
\end{equation}
\begin{equation}
IPD(t,k)=\angle\left(\frac{X_{\text{left}}(t,k)}{X_{\text{right}}(t,k)}\right)
\end{equation}
where $X_{left}(t,k)$ and ${X_{right}(t,k)}$ denote the left channel and right channel complex spectrum of the stereo audio. The operator $\left|\cdot\right|$ takes the absolute of the complex value, and $\angle(\cdot)$ finds the phase angle. In this work, we also define interaural magnitude differences (IMD) which is similar to the ILD. IMD is defined in linear domain while ILD is defined in logarithmic domain.
\begin{equation}
IMD(t,k)=\left|{X_{left}(t,k)}\right|-\left|{X_{right}(t,k)}\right|
\end{equation}
Fig. \ref{fig:cnn_imd} shows the structure of incorporating the spatial features (IMD/ILD/IPD, etc.) using an additional CNN. Then the activations learned from the basic features and the activations learned from the spatial features are concatenated to be fed into the GRU-RNN plotted in Fig. \ref{fig:cnn_rnn_at}.
\begin{figure}[t]
	\centering
	\centerline{\includegraphics[width=\columnwidth]{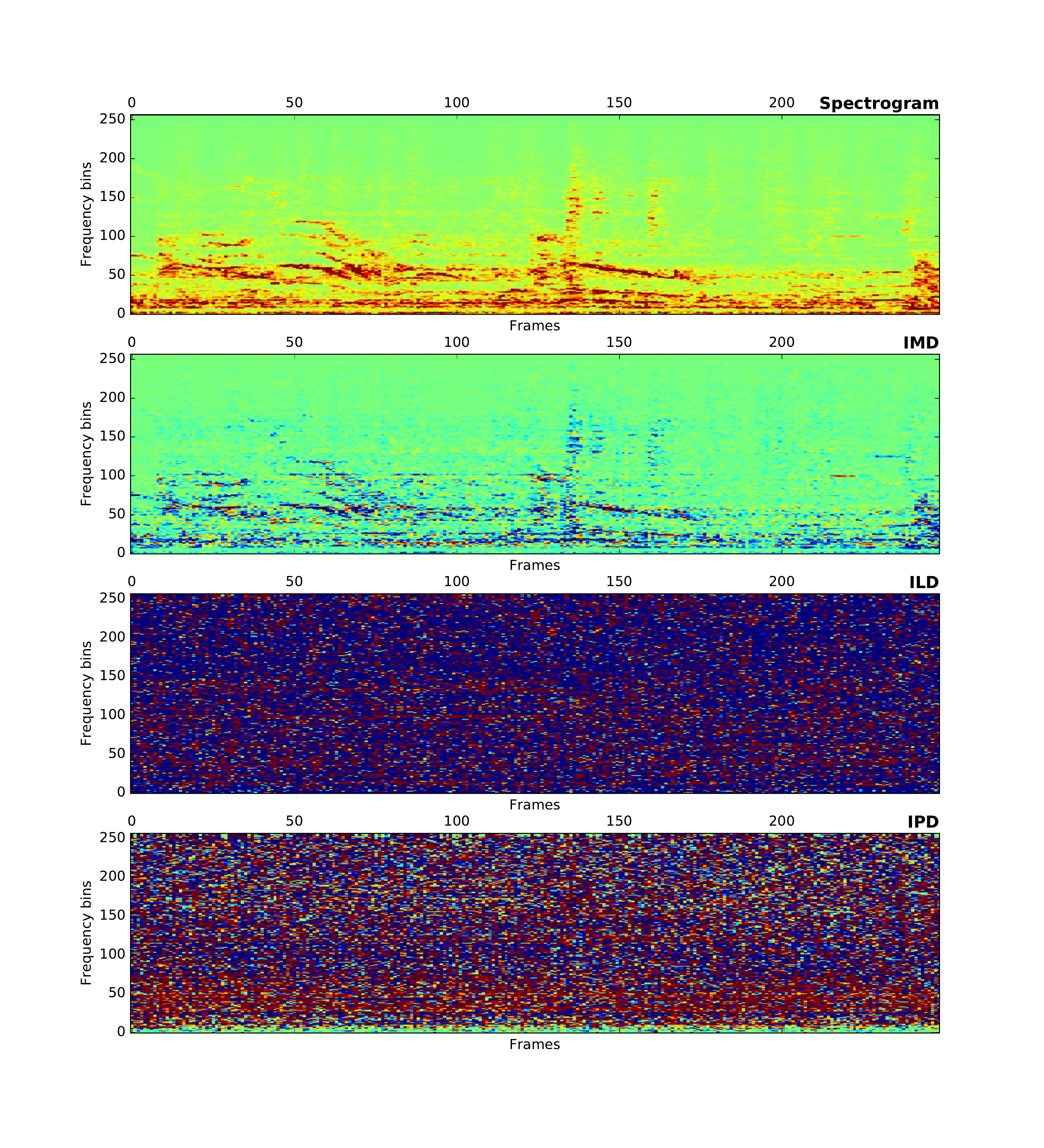}}
	\caption{The spectrogram, IMD, ILD and IPD of ac recording from the audio tagging task of DCASE 2016 challenge. The labels of this audio are ``child speech" and ``TV sound".}
	\label{fig:imd_ipd_ild_v2}
\end{figure}
The audio files of the audio tagging task of the DCASE 2016 challenge are recorded in a domestic home environment. There are severe reverberation, high-level background noise and multiple acoustic sources. These factors might influence the effectiveness of IPD and ILD. Fig. \ref{fig:imd_ipd_ild_v2} shows the spectrogram, IMD, ILD and IMD of one recording from the audio tagging task of DCASE 2016 challenge. The IMD appears to have some meaningful patterns while the ILD and the IPD seem to be random which would lead to the training difficulties of the classifier. From our empirical experiments, IPD and ILD do not appear to help to improve the classifier performance while IMD is beneficial. Similar results were reported in \cite{adavannesound} where IPD was found not to be helpful for the sound event detection in home scenes but helpful for the event detection of residential areas. This may be because residential areas are open areas with less reverberation than indoor home environments. Hence we will use IMD as our spatial features in this work. The filter size of CNNs learned on the IMD is set the same with the related configuration for the spectrogram.

\section{Data set and experimental setup}
\label{sec:exp_setup}
\subsection{DCASE 2016 audio tagging challenge}
\label{ssec:data_set}
We conducted our experiments based on the DCASE 2016 audio tagging challenge \cite{dcase_t4}. This audio tagging task consists of the five-fold development set and the evaluation set, which are built based on the CHiME-home dataset \cite{foster2015}. The audio recordings were made in a domestic environment \cite{christensen2010}. The audio data are provided as 4-second chunks at 48kHz sampling rates in stereo mode. We downsampled them into 16kHz sampling rate.


For each chunk, three annotators gave three labels, namely multi-label annotations. Then the discrepancies among annotators are reduced by conducting a majority vote. The annotations are based on a set of seven audio events as presented in Table \ref{tab:annotations}.
A detailed description of the annotation procedure is provided in \cite{foster2015}.
\begin{table}[h]
	\centering
	\caption{Seven audio events used as the reference labels.}
	\begin{tabular}{|c|c|}
		\hline
		audio event & Event descriptions \\ \hline
		`b' & Broadband noise \\ 
		`c' & Child speech \\
		`f' & Adult female speech \\
		`m' & Adult male speech \\
		`o' & Other identifiable sounds \\
		`p' & Percussive sound events, \\
		& e.g. footsteps, knock, crash \\
		`v' & TV sounds or Video games \\
		\hline
	\end{tabular}
	\label{tab:annotations}
\end{table}

\begin{table}[h]
	\centering
	\caption{The configurations for the five-fold development set and the final evaluation set of the DCASE 2016 audio tagging task.}
	\begin{tabular}{|c|c|c|}
		\hline
		Fold index of development set & \#Training chunks& \#Test chunks\\ \hline
		0 & 4004 & 383 \\ 
		1 & 3945 & 442 \\
		2 & 3942 & 463 \\
		3 & 4116 & 271 \\
		4 & 4000 & 387 \\
		\hline
		\hline
		\hline
		Evaluation set & 4387 & 816 \\
		\hline
	\end{tabular}
	\label{tab:eval_data}
\end{table}

\subsection{Experimental setup}\label{ssec:exp_setup}

In the experiments below, we follow the standard specification of the DCASE 2016 audio tagging task \cite{dcase_t4},
On the development set, we use the official five folds for cross-validation.
Table \ref{tab:eval_data} shows 
the number of chunks in the training and test set used for each fold. The number of final evaluation configuration is also listed.

The parameters of the networks are tuned based on the heuristic experience. All of the CNNs have 128 filters or feature maps. Following \cite{sainath2015learning}, the filter size for MFBs, spectrograms and raw waveforms are 30, 200, and 400, respectively. These parameters can form a large receptive field for each type of basic features considering that only one convolutional layer was designed. The CNN layer is followed by three RNN layers with 128 GRU blocks. One feed-forward layer with 500 ReLU units is finally connected to the 7 sigmoid output units. We pre-process each audio chunk by segmenting them using a $32$ms sliding window with a $16$ms hop size, and converting each segment into 40-dimension MFBs, 257-dimension spectrograms and 512-dimension raw waveforms. For performance evaluation, we use equal error rate (EER) as the main metric which is also suggested by the DCASE 2016 audio tagging challenge. EER is defined as the point of equal false negative ($FN$) rate and false positive ($FP$) rate \cite{murphy2012}. The source codes for this paper can be downloaded from Github\footnote{\url{https://github.com/yongxuUSTC/cnn_rnn_spatial_audio_tagging}}.

\subsection{Compared methods}
We compared our methods with the state-of-the-art systems. Lidy-CQT-CNN \cite{lidy2016cqt} and Cakir-MFCC-CNN \cite{cakirdomestic} won the first and the second prize of the DCASE2016 audio tagging challenge \cite{dcase_t4}. They both used convolutional neural networks (CNN) as the classifier. We also compare to our previous method \cite{xu2016fully} which demonstrated the state-of-the-art performance using de-noising auto-encoder (DAE) to learn robust features.

\section{Experimental results and analysis}
\label{sec:results}
In this section, the IMD effectiveness will be firstly evaluated on the development set of Task 4 of the DCASE 2016 challenge among the different features, i.e., spectrograms, MFBs and raw waveforms. Then the final evaluation will be presented by comparing with the state-of-the-art methods on the evaluation set of Task 4 of the DCASE 2016 challenge.
\subsection{The effectiveness of the IMD}
Table \ref{tab:imd_dev} shows the EER comparisons on seven labels among the spectrogram, the raw waveform and the MFB systems with or without the IMD information, which are evaluated on the development set of the DCASE 2016 audio tagging challenge. Firstly, we can compare the proposed convolutional gated recurrent neural networks on spectrograms, raw waveforms and MFBs. Spectrograms are better than the MFBs perhaps because the spectrogram has more detailed frequency information compared with the MFB. For example, spectrograms are much better than MFBs on child speech (denoted as `c') and female speech (denoted as `f') where a lot of high frequency information exists. The raw waveforms are worse than the spectrograms and the MFBs. One possible reason is that the learned FIR filters are not stable when the whole training set is small (about 3.5 hours of audio in this work). The same explanation was given in \cite{tuske2014acoustic} on the speech recognition task. \cite{sainath2015learning} shows that raw waveforms can get better recognition accuracy with the mel-spectra on 2000 hours Google voice search data.

With the help of the IMD spatial features, the EER are improved compared to all of the corresponding basic features alone. The raw waveforms with IMD can even get comparable results with the spectrograms and the MFBs. The MFB-IMD combination is slightly better than Spec-IMD, which may be because the IMD is calculated from the left and right spectrograms. The IMD has some common information with the spectrograms which can be seen from Fig. \ref{fig:imd_ipd_ild_v2}. However, the IMD is more complementary for the MFBs and the raw waveforms. The previous best performance on the development set of the DCASE 2016 audio tagging challenge was obtained in our recent work using denoising auto-encoder \cite{xu2016fully} with 0.126 EER, but here we get better performance with 0.10 EER.

\subsection{Overall evaluations}
Table \ref{tab:imd_eva} presents the EER comparisons on seven labels among Lidy-CQT-CNN \cite{lidy2016cqt}, Cakir-MFCC-CNN \cite{cakirdomestic}, our previous DAE-DNN \cite{xu2016fully}, and the proposed systems on the spectrogram, the raw waveform and the MFB systems with the IMD information, which are evaluated on the final evaluation set of the DCASE 2016 audio tagging challenge. The de-noising auto-encoder \cite{xu2016fully} was our recent work which can outperform the leading system in the DCASE 2016 audio tagging challenge, namely Lidy-CQT-CNN \cite{lidy2016cqt}. Our proposed convolutional gated recurrent neural network incorporating the IMD features in this work gives further improved performance. The MFB-IMD obtains the best performance with 0.123 EER which is the state-of-the-art performance on the evaluation set of the DCASE 2016 audio tagging challenge.

\begin{table}[t]
\scriptsize 
	\centering
	\caption{EER comparisons on seven labels among the spectrogram, the raw waveform and the MFB systems with or without the IMD information, which are evaluated on the\textbf{ development set} of the DCASE 2016 audio tagging challenge.}
\begin{tabular}{l|cccccccc}
	\hline
	Dev set& c     & m     & f     & v     & p     & b     & o     & ave \\
	\hline
	Spec  & 0.121 & 0.085 & 0.155 & 0.025 & 0.138 & 0.017 & 0.231 & 0.110 \\
	RAW   & 0.157 & 0.085 & 0.156 & 0.028 & 0.139 & 0.059 & 0.263 & 0.127 \\
	MFB   & 0.145 & 0.086 & 0.167 & 0.024 & 0.133 & 0.037 & 0.239 & 0.119 \\
	Spec-IMD & \textbf{0.120} & \textbf{0.080} & 0.143 & \textbf{0.012} & 0.115 & 0.023 & 0.232 & 0.104 \\
	RAW-IMD & 0.135 & 0.085 & 0.164 & 0.014 & \textbf{0.108} & \textbf{0.006} & 0.231 & 0.106 \\
	MFB-IMD & 0.125 & 0.086 & \textbf{0.140} & \textbf{0.012} & 0.110 & 0.011 & \textbf{0.230} & \textbf{0.102} \\
	\hline
\end{tabular}%
	\label{tab:imd_dev}
\end{table}

\begin{table}[t]
\scriptsize 
	\centering
	\caption{EER comparisons on seven labels among Lidy-CQT-CNN \cite{lidy2016cqt}, Cakir-MFCC-CNN \cite{cakirdomestic}, DAE-DNN \cite{xu2016fully}, and the proposed systems on the spectrogram, the raw waveform and the MFB systems with the IMD information, which are evaluated on the \textbf{final evaluation set} of the DCASE 2016 audio tagging challenge.}
\begin{tabular}{l|cccccccc}
	\hline
	Eval set& c     & m     & f     & v     & p     & b     & o     & ave \\
	\hline
	Cakir \cite{cakirdomestic} & 0.250 & 0.159 & 0.250 & 0.027 & 0.208 & 0.022 & 0.258 & 0.168 \\
	Lidy \cite{lidy2016cqt} & 0.210 & 0.182 & 0.214 & 0.035 & 0.168 & 0.032 & 0.320 & 0.166 \\
	DAE \cite{xu2016fully} & 0.210 & 0.149 & 0.207 & \textbf{0.022} & 0.175 & 0.014 & 0.256 & 0.148 \\
	RAW-IMD & 0.189 & 0.152 & 0.200 & 0.053 & 0.156 & \textbf{0.010} & \textbf{0.236} & 0.142 \\
	Spec-IMD & 0.166 & 0.165 & \textbf{0.143} & 0.024 & \textbf{0.123} & 0.034 & 0.250 & 0.129 \\
	MFB-IMD & \textbf{0.150} & \textbf{0.145} & \textbf{0.143} & 0.031 & 0.135 & 0.013 & 0.248 & \textbf{0.123} \\
	\hline
\end{tabular}%

	\label{tab:imd_eva}
\end{table}

\section{Conclusion}
\label{sec:conclusion}
In this paper, we propose a convolutional gated recurrent neural network (CGRNN) to learn on the mel-filter banks (MFBs), the spectrograms and even the raw waveforms. The spatial features, namely the interaural magnitude difference (IMDs), are incorporated into the framework and are demonstrated to be effective to further improve the performance. Spectrogram gives better performance than MFBs without the spatial features. However the MFBs with the IMDs can get the minimal EER, namely 0.102, on the development set of the DCASE 2016 audio tagging challenge. Raw waveforms give comparable performance on the development set. Finally, on the evaluation set of the DCASE 2016 audio tagging challenge, our proposed MFB-IMD system can get the state-of-the-art performance with 0.l23 EER. It is still interesting to further explore why the MFB-IMD system is better than the Spec-IMD system in our future work. In addition, we will also investigate the proposed framework to model raw waveforms on larger training datasets to learn more robust filters.


\section*{Acknowledgment}
This work was supported by the Engineering and Physical Sciences Research Council (EPSRC) of the UK under the grant EP/N014111/1. Qiuqiang Kong is partially supported by the China Scholarship Council (CSC).



%
\bibliographystyle{IEEEtran}
\bibliography{refs}

\begin{thebibliography}{10}
\providecommand{\url}[1]{#1}
\csname url@samestyle\endcsname
\providecommand{\newblock}{\relax}
\providecommand{\bibinfo}[2]{#2}
\providecommand{\BIBentrySTDinterwordspacing}{\spaceskip=0pt\relax}
\providecommand{\BIBentryALTinterwordstretchfactor}{4}
\providecommand{\BIBentryALTinterwordspacing}{\spaceskip=\fontdimen2\font plus
\BIBentryALTinterwordstretchfactor\fontdimen3\font minus
  \fontdimen4\font\relax}
\providecommand{\BIBforeignlanguage}[2]{{%
\expandafter\ifx\csname l@#1\endcsname\relax
\typeout{** WARNING: IEEEtran.bst: No hyphenation pattern has been}%
\typeout{** loaded for the language `#1'. Using the pattern for}%
\typeout{** the default language instead.}%
\else
\language=\csname l@#1\endcsname
\fi
#2}}
\providecommand{\BIBdecl}{\relax}
\BIBdecl

\bibitem{wold1996content}
E.~Wold, T.~Blum, D.~Keislar, and J.~Wheaten, ``Content-based classification,
  search, and retrieval of audio,'' \emph{IEEE {M}ultimedia}, vol.~3, no.~3,
  pp. 27--36, 1996.

\bibitem{giannoulis2013detection}
D.~Giannoulis, E.~Benetos, D.~Stowell, M.~Rossignol, M.~Lagrange, and M.~D.
  Plumbley, ``Detection and classification of acoustic scenes and events: an
  {IEEE AASP} challenge,'' in \emph{2013 IEEE Workshop on Applications of
  Signal Processing to Audio and Acoustics}, pp. 1--4.

\bibitem{cano2005content}
P.~Cano, M.~Koppenberger, and N.~Wack, ``Content-based music audio
  recommendation,'' in \emph{Proceedings of the 13th {A}nnual ACM
  {I}nternational {C}onference on {M}ultimedia}, 2005, pp. 211--212.

\bibitem{molau2001computing}
S.~Molau, M.~Pitz, R.~Schluter, and H.~Ney, ``Computing mel-frequency cepstral
  coefficients on the power spectrum,'' in \emph{Acoustics, Speech, and Signal
  Processing, 2001. Proceedings.(ICASSP'01). 2001 IEEE International Conference
  on}, vol.~1, pp. 73--76.

\bibitem{nadeu2001time}
C.~Nadeu, D.~Macho, and J.~Hernando, ``Time and frequency filtering of
  filter-bank energies for robust hmm speech recognition,'' \emph{Speech
  {C}ommunication}, vol.~34, no.~1, pp. 93--114, 2001.

\bibitem{kingsbury1998robust}
B.~E. Kingsbury, N.~Morgan, and S.~Greenberg, ``Robust speech recognition using
  the modulation spectrogram,'' \emph{Speech communication}, vol.~25, no.~1,
  pp. 117--132, 1998.

\bibitem{hinton2012deep}
G.~Hinton, L.~Deng, D.~Yu, G.~E. Dahl, A.-r. Mohamed, N.~Jaitly, A.~Senior,
  V.~Vanhoucke, P.~Nguyen, T.~N. Sainath \emph{et~al.}, ``Deep neural networks
  for acoustic modeling in speech recognition: The shared views of four
  research groups,'' \emph{IEEE Signal Processing Magazine}, vol.~29, no.~6,
  pp. 82--97, 2012.

\bibitem{xu2016fully}
Y.~Xu, Q.~Huang, W.~Wang, P.~Foster, S.~Sigtia, P.~J. Jackson, and M.~D.
  Plumbley, ``Unsupervised feature learning based on deep models for
  environmental audio tagging,'' \emph{arXiv preprint arXiv:1607.03681v2},
  2016.

\bibitem{espi2015exploiting}
M.~Espi, M.~Fujimoto, K.~Kinoshita, and T.~Nakatani, ``Exploiting
  spectro-temporal locality in deep learning based acoustic event detection,''
  \emph{EURASIP Journal on Audio, Speech, and Music Processing}, vol. 2015,
  no.~1, p.~1, 2015.

\bibitem{sainath2015learning}
T.~N. Sainath, R.~J. Weiss, A.~Senior, K.~W. Wilson, and O.~Vinyals, ``Learning
  the speech front-end with raw waveform {CLDNN}s,'' in \emph{Proc.
  Interspeech}, 2015.

\bibitem{tuske2014acoustic}
Z.~T{\"u}ske, P.~Golik, R.~Schl{\"u}ter, and H.~Ney, ``Acoustic modeling with
  deep neural networks using raw time signal for {LVCSR},'' in \emph{Proc. of
  Interspeech}, 2014, pp. 890--894.

\bibitem{golik2015convolutional}
P.~Golik, Z.~T{\"u}ske, R.~Schl{\"u}ter, and H.~Ney, ``Convolutional neural
  networks for acoustic modeling of raw time signal in lvcsr,'' in
  \emph{Sixteenth Annual Conference of the International Speech Communication
  Association}, 2015.

\bibitem{krizhevsky2012imagenet}
A.~Krizhevsky, I.~Sutskever, and G.~E. Hinton, ``Imagenet classification with
  deep convolutional neural networks,'' in \emph{Advances in {N}eural
  {I}nformation {P}rocessing {S}ystems}, 2012, pp. 1097--1105.

\bibitem{oord2016wavenet}
A.~v.~d. Oord, S.~Dieleman, H.~Zen, K.~Simonyan, O.~Vinyals, A.~Graves,
  N.~Kalchbrenner, A.~Senior, and K.~Kavukcuoglu, ``Wavenet: A generative model
  for raw audio,'' \emph{arXiv preprint arXiv:1609.03499}, 2016.

\bibitem{lidy2016cqt}
\BIBentryALTinterwordspacing
T.~Lidy and A.~Schindler, ``{CQT}-based convolutional neural networks for audio
  scene classification and domestic audio tagging,'' \emph{IEEE AASP Challenge
  on Detection and Classification of Acoustic Scenes and Events (DCASE 2016)}.
  [Online]. Available:
  \url{http://www.ifs.tuwien.ac.at/~schindler/pubs/DCASE2016b.pdf}
\BIBentrySTDinterwordspacing

\bibitem{cakirdomestic}
\BIBentryALTinterwordspacing
E.~Cak{\i}r, T.~Heittola, and T.~Virtanen, ``Domestic audio tagging with
  convolutional neural networks,'' \emph{IEEE AASP Challenge on Detection and
  Classification of Acoustic Scenes and Events (DCASE 2016)}. [Online].
  Available:
  \url{https://www.cs.tut.fi/sgn/arg/dcase2016/documents/challenge_technical_reports/Task4/Cakir_2016_task4.pdf}
\BIBentrySTDinterwordspacing

\bibitem{yundiscriminative}
\BIBentryALTinterwordspacing
S.~Yun, S.~Kim, S.~Moon, J.~Cho, and T.~Kim, ``Discriminative training of gmm
  parameters for audio scene classification and audio tagging,'' \emph{IEEE
  AASP Challenge on Detection and Classification of Acoustic Scenes and Events
  (DCASE 2016)}. [Online]. Available:
  \url{http://www.cs.tut.fi/sgn/arg/dcase2016/documents/challenge_technical_reports/Task4/Yun_2016_task4.pdf}
\BIBentrySTDinterwordspacing

\bibitem{eghbal2016cp}
H.~Eghbal-Zadeh, B.~Lehner, M.~Dorfer, and G.~Widmer, ``{CP-JKU} submissions
  for dcase-2016: A hybrid approach using binaural i-vectors and deep
  convolutional neural networks,'' Tech. Rep., DCASE2016 Challenge, Tech. Rep.,
  2016.

\bibitem{adavannesound}
\BIBentryALTinterwordspacing
S.~Adavanne, G.~Parascandolo, P.~Pertil{\"a}, T.~Heittola, and T.~Virtanen,
  ``Sound event detection in multichannel audio using spatial and harmonic
  features,'' \emph{IEEE Detection and Classification of Acoustic Scenes and
  Events workshop}, 2016. [Online]. Available:
  \url{http://www.cs.tut.fi/sgn/arg/dcase2016/documents/workshop/Adavanne-DCASE2016workshop.pdf}
\BIBentrySTDinterwordspacing

\bibitem{ciresan2011flexible}
D.~C. Ciresan, U.~Meier, J.~Masci, L.~Maria~Gambardella, and J.~Schmidhuber,
  ``Flexible, high performance convolutional neural networks for image
  classification,'' in \emph{{IJCAI} Proceedings-International Joint Conference
  on Artificial Intelligence}, vol.~22, no.~1, 2011, p. 1237.

\bibitem{mikolov2010recurrent}
T.~Mikolov, M.~Karafi{\'a}t, L.~Burget, J.~Cernock{\`y}, and S.~Khudanpur,
  ``Recurrent neural network based language model,'' in \emph{Proc.
  Interspeech}, 2010.

\bibitem{graves2013speech}
A.~Graves, A.-R. Mohamed, and G.~Hinton, ``Speech recognition with deep
  recurrent neural networks,'' in \emph{2013 IEEE international conference on
  acoustics, speech and signal processing}, 2013, pp. 6645--6649.

\bibitem{lin2013network}
M.~Lin, Q.~Chen, and S.~Yan, ``Network in network,'' \emph{arXiv preprint
  arXiv:1312.4400}, 2013.

\bibitem{abdel2013exploring}
O.~Abdel-Hamid, L.~Deng, and D.~Yu, ``Exploring convolutional neural network
  structures and optimization techniques for speech recognition.'' in
  \emph{Proc. Interspeech}, 2013, pp. 3366--3370.

\bibitem{pascanu2013difficulty}
R.~Pascanu, T.~Mikolov, and Y.~Bengio, ``On the difficulty of training
  recurrent neural networks.'' \emph{ICML}, pp. 1310--1318, 2013.

\bibitem{hochreiter1997long}
S.~Hochreiter and J.~Schmidhuber, ``Long short-term memory,'' \emph{Neural
  {C}omputation}, vol.~9, no.~8, pp. 1735--1780, 1997.

\bibitem{wu2016investigating}
Z.~Wu and S.~King, ``Investigating gated recurrent networks for speech
  synthesis,'' in \emph{2016 IEEE International Conference on Acoustics, Speech
  and Signal Processing (ICASSP)}, pp. 5140--5144.

\bibitem{cho2014learning}
K.~Cho, B.~Van~Merri{\"e}nboer, C.~Gulcehre, D.~Bahdanau, F.~Bougares,
  H.~Schwenk, and Y.~Bengio, ``Learning phrase representations using rnn
  encoder-decoder for statistical machine translation,'' \emph{arXiv preprint
  arXiv:1406.1078}, 2014.

\bibitem{chung2014empirical}
J.~Chung, C.~Gulcehre, K.~Cho, and Y.~Bengio, ``Empirical evaluation of gated
  recurrent neural networks on sequence modeling,'' \emph{arXiv preprint
  arXiv:1412.3555}, 2014.

\bibitem{kingma2014adam}
D.~Kingma and J.~Ba, ``Adam: A method for stochastic optimization,''
  \emph{arXiv preprint arXiv:1412.6980}, 2014.

\bibitem{blauert1997spatial}
J.~Blauert, \emph{Spatial {H}earing: the {P}sychophysics of {H}uman {S}ound
  {L}ocalization}.\hskip 1em plus 0.5em minus 0.4em\relax MIT press, 1997.

\bibitem{dcase_t4}
\url{http://www.cs.tut.fi/sgn/arg/dcase2016/task-audio-tagging}.

\bibitem{foster2015}
P.~Foster, S.~Sigtia, S.~Krstulovic, J.~Barker, and M.~Plumbley,
  ``{CHiME}-home: A dataset for sound source recognition in a domestic
  environment,'' in \emph{IEEE Workshop on Applications of Signal Processing to
  Audio and Acoustics}, 2015, pp. 1--5.

\bibitem{christensen2010}
H.~Christensen, J.~Barker, N.~Ma, and P.~Green, ``The {CHiME} corpus: a
  resource and a challenge for computational hearing in multisource
  environments,'' in \emph{Proceedings of Interspeech}, 2010, pp. 1918--1921.

\bibitem{murphy2012}
K.~P. Murohy, \emph{Machine Learning: A Probabilistic Perspective}.\hskip 1em
  plus 0.5em minus 0.4em\relax MIT Press, 2012.

\end{thebibliography}


\end{document}